\documentclass[11pt]{article}
\usepackage[table,xcdraw]{xcolor}
\usepackage[utf8]{inputenc}
\usepackage[margin=2cm]{geometry}
\usepackage{graphicx}
\usepackage{setspace}
\usepackage{amsmath}
\usepackage{amssymb}
\usepackage{url}
\usepackage{hyperref}
\usepackage{xcolor}
\usepackage{tikz}
\usepackage{xcolor}
\usepackage{tabularx}
\usepackage{titlesec}
\usepackage[defaultlines=2,all]{nowidow}
\usepackage{caption}
\titlespacing{\paragraph}{0pt}{0pt}{1ex}

\usepackage[margin=2cm,font=footnotesize]{caption}

\usepackage{rotating}
\usepackage{tikz}

\usepackage[backend=biber,style=authoryear,
citestyle=authoryear,isbn=false,url=false,eprint=false,
giveninits=true,maxcitenames=1,uniquename=init]{biblatex}
\AtEveryBibitem{
  \clearfield{note}
}

\definecolor{bleucite}{RGB}{34,111,212}
\usepackage{hyperref}
\hypersetup{hidelinks}
\AtEveryCite{\color{bleucite}}

\graphicspath{ {./images/} }
\newcommand{\ts}{\textsuperscript}

\newcommand{\beginsupplement}{%
        \setcounter{table}{0}
        \renewcommand{\thetable}{S\arabic{table}}%
        \setcounter{figure}{0}
        \renewcommand{\thefigure}{S\arabic{figure}}%
     }

\title{Socio-economic Segregation in a Population-Scale Social Network 
}

\author{Yuliia Kazmina$^{1,*}$, Eelke M. Heemskerk$^1$, Eszter Bokányi$^1$, Frank W. Takes$^2$}

\date{{\footnotesize%
    $^1$University of Amsterdam\\
    $^2$Leiden University\\
    $^*$\href{mailto:by.kazmina@uva.nl}{y.kazmina@uva.nl}\\[2ex]
    \textit{version as of 22nd September 2023}
}}

\providecommand{\keywords}[1]{\textbf{\textit{Keywords: }} #1}

\begin{document}

\onehalfspacing

\maketitle

\abstract{

We propose a social network-aware approach to study socio-economic segregation. The key question that we address is whether patterns of segregation are more pronounced in social networks than in the common spatial neighborhood-focused manifestations of segregation. We, therefore, conduct a population-scale social network analysis to study socio-economic segregation at a comprehensive and highly granular social network level. For this, we utilize social network data from Statistics Netherlands on 17.2 million registered residents of the Netherlands that are connected through around 1.3 billion ties distributed over five distinct tie types. 
We take income assortativity as a measure of socio-economic segregation, compare a social network and spatial neighborhood approach,  and find that the social network structure exhibits two times as much segregation. As such, this work complements the spatial perspective on segregation in both literature and policymaking. While at a widely used unit of spatial aggregation (e.g., the geographical neighborhood), patterns of socio-economic segregation may appear relatively minimal,  they may in fact persist in the underlying social network structure. Furthermore, we discover higher socio-economic segregation in larger cities, shedding a different light on the common view of cities as hubs for diverse socio-economic mixing. A population-scale social network perspective hence offers a way to uncover hitherto “hidden” segregation that extends beyond spatial neighborhoods and infiltrates multiple aspects of human life.}

\setlength{\parindent}{0em}
\setlength{\parskip}{0.8em}


\section{Introduction}

Social segregation is one of the most persistent problems that stand in the way of upward socio-economic mobility. It is typically viewed as the degree to which two or more groups live separately from one another \parencite{massey1988}, voluntarily, as a result of policies, or due to a reinforcing combination of both. As such, segregation seeps into every facet of life. The common approach to study segregation is to analyze the spatial distribution of groups of people with different characteristics, zooming in on neighborhoods. Indeed, neighborhoods are to a large extent representative of early schooling opportunities, provide access to a particular set of neighborhood relations, and create other opportunities for people to interact. They are also relevant because the probability of establishing social ties decreases with geographical distance \parencite{Liben-Nowell2005, Small2019, Viry2012, Bidart22}. Hence, a neighborhood’s social composition to a certain extent defines its residents’ social and economic opportunities \parencite{Atkinson2001, Chetty2018, Friedrichs2003, Mayer1989}. 

However, the all-pervasive nature of segregation extends well beyond neighborhoods as the limited exposure of various population groups to one another is inherent to all facets of life including social circles, lifestyle practices, education trajectories, or career choices \parencite{Bojanowski2014, Henry2011, Hofstra2017}. High levels of segregation therefore lead to the emergence of “socio-economic bubbles” where people almost exclusively interact with similar others (e.g. the \emph{filter bubble} concept, see \cite{eli2011filter}). Hence, it is not obvious that administrative neighborhoods are the best level of analysis to measure, study, and combat segregation. First of all, individuals’ social networks are constructed by a wide variety of social contexts that may well extend beyond the spatially defined neighborhood, such as family, school classes, or colleagues. Second, some convincingly argue that neighborhoods in postindustrial societies have lost much of their previous importance for providing social cohesion and as the locus for social activities \parencite{Dahlin2008, bookDuyvendak, Pinkster2014}. Thus, notwithstanding the important contributions made by the work on residential segregation, the spatial manifestation of social segregation in neighborhoods remains a smaller share of a wider issue \parencite{Leo2016, Dong2020, Bidart22, Morales2019}. 

We develop a social network-aware approach to study socio-economic segregation and ask the question of whether patterns of segregation in social networks are more pronounced than the common spatial manifestations of segregation. We intend to uncover potential “hidden segregation” that is not captured by spatial neighborhoods. In addition, we want to study segregation as a multi-layered phenomenon in the sense that it plays out across different sets of social contexts. To capture this, we conduct a population-scale social network analysis \parencite{Bokanyi2022} to analyze socio-economic segregation at a comprehensive and highly granular level. Recent advances in data collection have made it possible to investigate large-scale social networks, for instance using digital traces of online platforms \parencite{Chetty2022, Chetty2022a} as well as high-quality register data collected by national statistical bureaus \parencite{Ludvigsson2016, Erlangsen2015}. We build on the latter as a promising opportunity to study social cohesion and segregation and investigate individual-level information on various social ties of an entire country’s population: around 17.2 million registered residents of the Netherlands as of October 2018 linked through around 1.3 billion ties living in around 7.7 million households that are connected through around 914 million ties \parencite{laan_emery_2021}. With these data created by Statistics Netherlands, it is possible to represent the entire population as a large-scale social network in which nodes are individuals or households, and the relationships between them are based on the relations defined in the state-administered registers.

Population-scale network analysis has considerable benefits over existing approaches to social network inference \parencite{Peel2022} that rely on a wide variety of data collection designs ranging from study-specific surveys and name or position generators to larger-scale online social network data and mobile communication networks. Surveys for instance can be highly tailored to specific analytical goals, but are prone to sampling-related issues and non-response or cognitive bias. Moreover, they are costly to implement for larger sample sizes \parencite{Bruggen2011, Bruggen2016, Bosnjak2005}. Online social network data such as Meta/Facebook or Twitter do offer a cost-effective solution to obtain large amounts of social network data of self-reported social ties of individuals. However, these sources still suffer from issues of sample representativeness as well as the absence of information on the different types of relationships between individuals and individual node-level characteristics \parencite{Bokanyi2022, Lazer2021}. In a similar way, while mobile phone communication records capture particular social ties between individuals and are also well suited for measuring the intensity and temporal frequency of interactions at a granular level, they nevertheless leave no room for distinguishing different social contexts \parencite{Onnela2007, Eagle2009}. As such, it is hard to study the multi-layered structure of social segregation through such data. 

Official register data as used in this work offer, in contrast, a clear and precise definition for each type of social tie such as a household member, a classmate, or a colleague. In addition, it also comes with a wide range of high-quality attribute data from those same registers. These register-based population-scale social network data thus allow us to study various different patterns of social segregation without a priori spatial assumptions.  As opposed to more informal ties such as retweets, phone calls, or friendships, the social ties encoded in official registers are so-called \emph{formal ties}, i.e. officially recognized or institutionalized ties and affiliations of residents such as their family connections, next-door neighbors, household members, school classmates or colleagues. It has been shown that formal ties capture the majority of people’s strong connections \parencite{Wrzus2013, BUIJS202225, VanEijk2010}. We consider the population-scale social network data to be a comprehensive mapping of the \emph{social opportunity structure} available to individuals \parencite{Bokanyi2022}. However, we also acknowledge the limitations the chosen approach suffers from. Administrative registers that serve as an input for the construction of the population-scale social network do not offer a way of quantifying or distinguishing the strength of a tie and intensity of communication between individuals. For instance, we know that two students belong to the same class, nevertheless, we are not aware if they communicate at all, what the nature of their relationship is (unless they share a tie in any other social context covered by the data), or if it is a positive or a negative tie, etc. Furthermore, the chosen approach has a limited potential for tailoring the input data to a specific research question. We rely on multiple existing registers being collected for various administrative purposes and while it is not impossible to modify registers’ design, such a task is a long-term policy initiative that would involve multiple stakeholders.

In the next section, we first summarize the literature on social segregation and compare the spatial neighborhood approaches with a broader social network approach, followed by an introduction of the Netherlands as our site of research. The subsequent section discusses the population-scale social network data as well as the methods used in the analysis. The empirical section presents the results of that analysis and shows on average doubled levels of social segregation in the social network approach compared to a spatial neighborhood approach. Further decomposition of the social network segregation patterns by layers (social contexts) as well as by various regions reveals even higher levels of segregation for particular types of social ties and subgroups of the population. In the discussion and conclusion, we reflect on the consequences of this previously “hidden” social segregation for policy interventions as well as on the opportunities of population-scale social network analysis as an efficient approach to map and analyze social segregation. 

\section{A social network perspective on socio-economic segregation}

The post-war trend of economic growth in the second half of the 20th century has benefited many but also led to a rather heterogeneous distribution of wealth among different socio-economic groups \parencite{Stiglitz2015, DablaNorris2015}. This is at odds with the central prediction of the neoclassical growth theory that envisions conditional convergence of disparities between countries, regions, neighborhoods, and households. Despite state and subnational efforts to tackle these issues both by implementing equality-enhancing policies and by increasing market forces, socio-economic inequalities remain self-reinforcing “traps” or “low-level equilibria” that are remarkably persistent over time \parencite{RobertMoffitt2018}. There are many explanations suggested for the long-term persistence of social divisions and related inequalities such as the intergenerational transmission of wealth and ability; capital market imperfections imposing credit constraints for mobility; local spatial segregation patterns; and self-fulfilling beliefs \parencite{Piketty2000}. However, the root cause of segregation may lay deeper in the social fabric of our societies \parencite{Chang2020}. We know that social networks do not emerge randomly. Far from it, social networks are typically segregated structures, where people share ties with similar others with respect to a variety of characteristics.  This is in part due to homophily: there is a higher probability for an individual to share a social tie with an individual with similar attributes such as, for instance, age, gender, language, nationality, religion \parencite{Melamed2020, McPherson2001}, or socio-economic-status \parencite{Leo2016, Morales2019, Dong2020, Bokanyi2021}. When the well-endowed cluster together, and those in a more disadvantaged position do so as well, these social cleavages make existing inequalities persistent. They may even amplify them \parencite{MaartenvanHamTiitTammaru2018} because social ties provide a form of capital as an infrastructure towards certain resources \parencite{Bourdieu1986}. Social capital literature firmly establishes that social relations provide much-needed access to a wide range of resources that are important for reducing inequalities and upward mobility. The resources that may flow through social networks include social and financial opportunities \parencite{Uzzi1999}, valuable information relevant for job search \parencite{Demchenko2011, Granovetter1973, Montgomery1992, Rajkumar2022} and educational opportunities \parencite{Frank2018}. Using diverse resources embedded in one's social network results in better socio-economic outcomes \parencite{Lin1982} and access to such resources is in turn determined by “the strength of [social] position” \parencite{Campbell1986, Lin1999}. 

Following a long tradition of social capital research, recent studies that have access to large-scale social network data empirically demonstrated how social networks among others explain socio-economic inequality \parencite{Jackson2010, Chetty2022, Chetty2022a}. Jackson finds that “immobility results when people end up trapped by the social circumstances into which they are born: the networks in which they are embedded fail to provide them with the information and opportunities that they need to succeed” (\cite{Jackson2020}, p. 118). And other work shows how “stronger” personal social networks characterized by larger quantities of “high-quality” social contacts are associated with socio-economic benefits \parencite{Pinquart2000, Woolcock2001, Chetty2022, Chetty2022a}. The social fabric people are embedded in is therefore best seen as a social opportunity structure \parencite{Bokanyi2022}. Roberts \parencite{Roberts1977, Roberts2009} theorized how a complex interplay of both external and internal sociological factors (such as family background, school choices, and labor market conditions) limits the choices available to an individual. He showed for instance that macro-level labor market imbalances, particularly youth unemployment, stem not from exclusively individual “wrong choices” but are driven and reinforced by the existing opportunity structures \parencite{Roberts2009}. In what follows, we build upon this conceptualisation of the social fabric and social capital embedded in it as a social opportunity structure. The extent to which this opportunity structure is segregated into different socio-economic foci or bubbles creates barriers to the accumulation of social capital. Socio-economic segregation prevents individuals from mixing with a heterogeneous set of people which would ensure the creation of a diverse profile of various types of social capital that would help tackle pervasive inequalities in a society. 

Segregation of social opportunity structure as conceptualized in this study presents the output homophily, i.e. the level to which social opportunities are segregated from the socio-economic perspective. While there is a complex interplay of the generative mechanisms behind the segregation outcomes we observe, our goal in this study is to compare the extent to which the population-scale social network is a segregated structure vs the common indicators of segregation on the level of administrative neighborhoods. The extensive body of literature on segregation offers a wide variety of contributing factors and structural properties that drive the segregation outcomes we observe. Among these are induced homophily (segregation outcomes driven by structural properties), choice homophily (indicative of the strength of personal preference to mix with similar others) as well as their complex interplay \parencite{firmansyah_pratama_2021}. Structural features include the distribution of attributes, relative size of minority groups as well as network topology \parencite{peel2020,karimi2023inadequacy} --- all these parameters have implications for the bounds of segregation level. They are further impacted by the correlation of multiple attributes and the composition of various social contexts within the population. All these different factors could possibly contribute to the output homophily we are able to capture, however, establishing causal claims on what drives segregation and to what extent is out of the scope of this study.

The relation between social segregation and inequalities has long been recognised by policymakers and many efforts have therefore been made to combat segregation. Most attention has been given to the spatial manifestation of social segregation. This is because segregation has been historically referred to in the context of deliberate policies that aim to separate or isolate specific groups of people within neighborhoods, cities, or regions. However, such an approach does not extend to the social and economic implications of such spatial divisions.
On the one hand, geographic neighborhoods are indeed directly related to residents' social and economic opportunities, as a fair share of social life plays out inside neighborhoods \parencite{Lenahan2022, Kuyvenhoven2021}. They provide a site of access to a particular set of neighbors and in general, create opportunities for people to interact in their community centers, sports clubs, and other civic society organizations \parencite{Friedrichs2003, Sharp2018}. On the other hand, spatial planning also requires strategic reasoning about desired neighborhood composition, because that is the “level of analysis” of urban development. Together, this underpins the dominance of a spatial perspective on segregation in both literature and policymaking. For instance, smart urban planning approaches see cities as the “champions of diversity” so that they provide larger sets of social choices which results in lower levels of segregation \parencite{Jacobs1961, Milgram1970, Glaeser2011, Bettencourt2013}. The policy focus also triggered a large body of literature on neighborhood effects and the “moving to opportunity” kind of social experiments. In these experiments, families were moved to other, more well-off, neighborhoods, and this change in their residential environment has been reported to improve intergenerational socio-economic-mobility, long-term health, education, and employment outcomes \parencite{Chetty2016, Chetty2018, Ludwig2013, Orr2006}. 

There are, however, good reasons to assume that the importance of neighborhoods as key social foci is diminishing. First, increased geographic mobility further adds to the spatial dispersion of social connections, discounting the importance of neighborhoods \parencite{Liben-Nowell2005, Viry2012, Bokanyi2021}. Second, there is a diminishing role for the neighborhood as a site for social tie (re)production. Already, co-worker ties are much more frequent than those to neighbors \parencite{Dahlin2008}. Third, such spatial dispersion is exacerbated by the rise of online social networks, which further reduce the importance of spatial residential patterns for social cohesion and segregation \parencite{Scellato2021}. In sum, while the spatial focus on neighborhoods in both literature and policymaking is understandable, it may underestimate actual patterns of social segregation. We therefore follow a population-scale network analysis approach to map and analyze the patterns of segregation for an entire country across various social contexts \parencite{laan_emery_2021}. The unit of analysis will be the household because these integrate their members into joint social and economic units significantly sharing their life trajectories, activities, and responsibilities. Individual-level income estimates might provide incomplete information on one’s socio-economic standing if they are integrated into a multiple-person household. A child or a stay-at-home parent might have zero income reported, and still be a part of a high socio-economic status household. Some of the consumption and lifestyle choices relevant to the socio-economic status are made on the level of a household, such as real estate decisions (with mortgage or rental costs), educational choices, etc. Households are therefore widely regarded as the locus for socio-economic status as well as the main target for welfare policies \parencite{Katz2001, Kling2007}. Households are also a basic unit of statistical data collection which provides a rich source of relevant socio-economic information on that level. 

We choose to investigate the social opportunity structure of the relatively egalitarian country of the Netherlands. Despite ranking high in terms of social cohesion as well as being in the top-tier of high-income economies, the Netherlands has been reported to have one of the highest wealth inequalities in the world as measured by the Gini coefficient \parencite{Suisse2019}. Wealth distribution in the Netherlands is heavily skewed as it is largely influenced by housing and pension practices that lead to more than half of the population having none to negative net wealth \parencite{Salverda2013}. On the other hand, income distribution is more evenly spread. As ranked by the income inequality expressed by the Gini coefficient, the Netherlands has consistently been reported in the top fifteen best-performing economies \parencite{Salverda2013}. The combination of relatively high wealth concentration and moderate income inequality makes the Netherlands a prime case study for social network segregation patterns that persist despite strong welfare state efforts. The state-administered data collection effort in the country also provides a unique and innovative way of constructing a large-scale social network data relevant to the study of segregation and inequalities.  

\section{Data and methods}
\label{sec:methods}

\textbf{Data.} The social network as well as corresponding node attributes are sourced from the Dutch population registers curated by Statistics Netherlands (Centraal Bureau voor de Statistiek, CBS). 
Using these administrative data, CBS has built a network covering the entire Dutch population \parencite{laan_emery_2021}. 
These network data contain information on 17.2 million registered residents of the Netherlands as of October 2018 that are linked through 1.3 billion ties. These ties can be grouped into five main types: family, household, colleagues, next-door neighbors, and classmates. The dataset is pseudonymized before research access, and it does not contain personal information such as address or date of birth. Because of the sensitive nature of the information, storage, and analysis have been conducted within the secure server environments of Statistics Netherlands. Further details about data generation, ensuring data security and privacy are described in \parencite{VanderLaan2021}.

\begin{figure}[h]
\centering
\includegraphics[width=0.5\textwidth,scale=0.05]{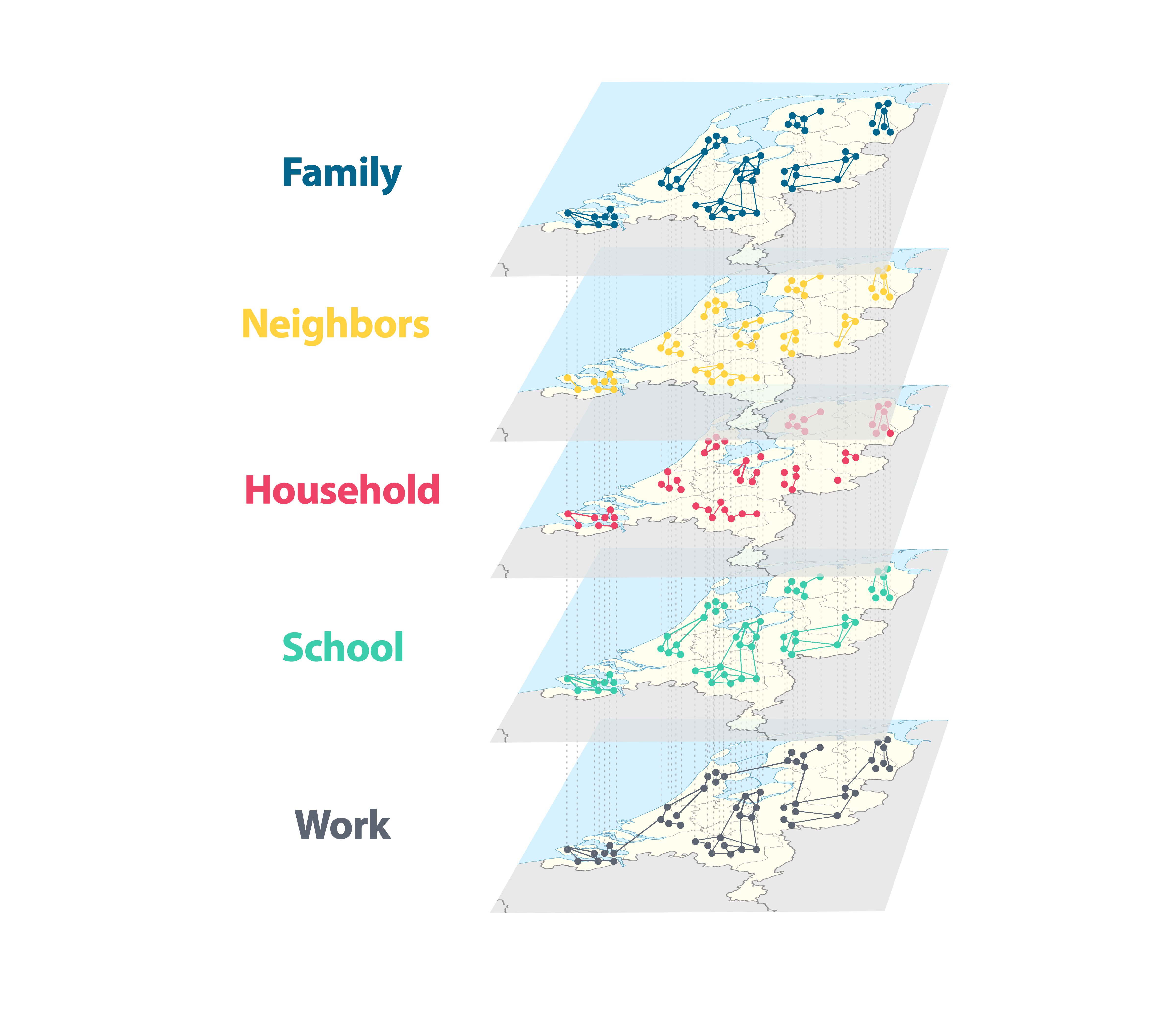}
\caption{\textbf{Schematic representation of the multilayer population-scale social network}: nodes are registered residents of the Netherlands, and links are ties in one or more social contexts: family, next-door neighbors, household members, school classmates, and colleagues.}
\label{fig:network_rep}
\end{figure}

\textbf{Network construction.} The schematic representation of the input CBS population-scale persons' network is presented in Figure~\ref{fig:network_rep}. Based on this network of people, we first perform a household-level aggregation of the network such that the result is a multilayer network in which nodes are the households and connections between them are family, neighborhood, school, and work ties that connect any of the members of households. The network is unweighted. Whenever households share multiple ties in a social context, a single link per layer is preserved. For instance, if parents in two households work together, and their kids are in the same school class, respective households share two links: one work link as well as one school tie. 
Except for single-person households, the majority of the households are either nuclear or extended family members living at the same address. Households not covered by kinship or institutional relations are classified as “other households”, consisting of individuals that live at the same address and act as a joint economic unit as defined by Statistics Netherlands \parencite{Witvliet2012}. For the purposes of this study, we exclude institutional households such as care homes, prisons, orphanages, etc. from the analysis. Table~\ref{tab:network_properties} summarizes properties of the transformed network providing information on the number of nodes and edges along with the average, minimum, and maximum degrees for each configuration. 

\begin{table}[h]
\centering
\begin{tabular}{lrrrrr}
\hline
\textbf{Network layer combination}                       & \textbf{Nodes}   & \textbf{Edges}  & \textbf{\textless{}k\textgreater{}} & \textbf{k\textsubscript{min}} & \textbf{k\textsubscript{max}} \\
\hline
Persons network \emph{P} (all layers)               & 17.2 M & 1.3 B & -                                         & -          & -              \\
Households network \emph{HH} (all layers)            & 7.7 M  & 914 M & 118                                       & 1          & 2 153          \\
- Households network (Family layer \emph{F})               & 7.7 M  & 132 M      & 18                                        & 1          & 265            \\
- Households network (School layer \emph{S})                & 2.2 M  & 208 M      & 96                                        & 1          & 1 706           \\
- Households network (Work layer \emph{W})                  & 4.6 M  & 504 M      & 110                                       & 1          & 1 706           \\
- Households network (Next-door neighbors layer \emph{N}) & 7.6 M  & 70 M      & 9                                         & 1          & 28  \\          
\hline

\end{tabular}
\caption{\textbf{Network properties for various network layer combinations.} 
Columns show the number of non-isolated nodes, edges,  average degree \textless{}k\textgreater{}, minimum degree k\textsubscript{min} and maximum degree k\textsubscript{max}.}
\label{tab:network_properties}
\end{table}

\textbf{Node attributes.} Based on the individual-level node attributes of household members, we assign joint characteristics of a household with respect to income, educational level, ethnic and migration background. Joint household-level attributes are derived from the profile of “founding” adults: either one adult in the household in case of single-person or single-parent households or based on the “founding” couple in other cases. If a founding adult in a household is single, attributes with respect to the adult's income, education, ethnic, and migrant groups are assigned to a corresponding household. For households with a founding couple, we combine individual characteristics of founding adults from a couple into a joint household-level attribute. In such cases, continuous variables such as income correspond to an equivalised combined household income, i.e. a value that takes into account the size and composition of a household. For nominal data that include ethnic group and migrant generation, we combine the categories describing individuals as provided by Statistics Netherlands. A list of categories that the ethnic group variable takes is shown in Table~\ref{tab:attributes_properties}\footnote{Since 2022, CBS introduced a new definition of migrant background (\url{https://www.cbs.nl/en-gb/news/2022/07/cbs-introducing-new-population-classification-by-origin}). This paper is based on the previous definition.}. The ethnic group variable is derived based on the country of origin of an individual and their parents. For the migrant generation, we distinguish between native Dutch, first- and second-generation migrants. Given that two individuals belong to the same ethnic group and/or migrant generation, household-level characteristic takes the identical value. Whenever two founding adults come from different ethnic or migrant generation backgrounds, we combine their individual-level characteristics into a mixed household’s attribute. For instance, a native Dutch living with a Surinamese would be described as a mixed Dutch-Surinamese household. For ordinal data describing people’s highest achieved education level, we follow the same aggregation approach. On the individual level education background takes a natural ordering from primary education as detailed in Table \ref{tab:attributes_properties}. When aggregated on a household level, the education attribute takes a corresponding value of the individuals if they obtained the same education level or presents a mixed category, for instance, MBO-WO (vocational and higher education). Information about node attributes is summarized in Table \ref{tab:attributes_properties}, and their distributions are presented in Supplementary Figure ~\ref{fig:hh_distributions_combined}. Correlations between household-level attributes are examined in Supplementary Table ~\ref{tab:attr_correlation}.

\small
\begin{table}[h]
\centering
\begin{tabular}{llll}
\hline
\textbf{Node attribute}     & \textbf{Type}       & \textbf{Possible values}          \\ \hline
\emph{Income}   & Continuous & N/A    
\\
\hline
\emph{Education}          & Ordinal    & \begin{tabular}[c]{@{}l@{}} Primary \\ VMBO (secondary applied)  \\ HAVO/VWO (secondary scientific) \\ MBO (vocational)   \\ WO (higher)

\end{tabular}                             \\
\hline
\emph{Migrant generation} & Nominal    & \begin{tabular}[c]{@{}l@{}}Native\\ First generation\\ Second generation\end{tabular}                                                                                                                                                                                                                                                                                                                                                                     \\
\hline
\emph{Ethnic group}       & Nominal    & \begin{tabular}[c]{@{}l@{}}Dutch\\ Moroccan\\ Turkish\\ Surinamese\\ Former Netherlands Antilles and   Aruba\\ Other non-western\\ Other western\end{tabular}                                                                                                                                                                                                                                                                         \\ 
\hline
\end{tabular}
\caption{\textbf{Node attributes}}
\label{tab:attributes_properties}
\end{table}

\textbf{Network layers.} We distinguish four network layers. First is the family layer which is based on child-parent information and allows us to derive links including an individual's parents, children, siblings, grandparents/grandchildren as well as aunts/uncles and cousins. Second is the school layer which is based on people who are admitted to a formal institute for education at the time of the data collection (October 2018). The layer distinguishes between primary, secondary/specialized secondary, vocational, and higher education levels. For all the school levels, people are grouped in classes if they attend the same educational institution, at the same location (applicable to elementary, secondary/secondary special, and higher education), with an equal overall duration of schooling and identical type of education (that can be distinguished among secondary, vocational, and higher educational institutions). Third is the work layer which provides a set of social ties between colleagues who work for the same employer. In the case of larger workplaces, only the 100 geographically closest colleagues (in terms of their place of residence) are sampled. Finally, the neighbor layer includes at most 10 geographically closest neighboring households, referred to as “next-door neighbors”. We include this as part of the social opportunity structure derived from the context of the built environment. Except for family, all layers in our study have a spatial angle --- next-door neighbors are defined strictly spatially, but lower levels of education, especially primary schooling, or workplace choice are also to an extent influenced by spatial constraints. However, these layers are not restricted by an a priori spatial unit of aggregation. Geographically, such social opportunity structure can span large geographical distances, yet social ties included in such analysis are one step away in terms of social network distance. Predefined spatial delineation is the case in a strict neighborhood approach dominant in the literature that we compare with. 

\textbf{Segregation measurement.} In what follows, we outline how we compare the social network perspective on segregation with a neighborhood perspective. The concept of a  neighborhood here denotes people living in the same geographic neighborhood as defined by the smallest administrative neighborhood borders in the Netherlands (the so-called “buurt” in Dutch), typically consisting of a few thousand households. The social network perspective is tailored to the considered population-scale multilayer network consisting of family, classmates, colleagues, and next-door neighbors explained above. To measure segregation from the spatial and network perspective, we take a three-step approach as detailed below. 

\emph{Grouping households.} We first categorize households into respective groups as defined by their attributes: education level, ethnic group, migrant generation, and household income. For nominal data (education, ethnic and migration background) the size of each group corresponds to the prevalence of each attribute in the population of households. For the continuous variable in the analysis (income) we perform a normalization. We split the full income range of Dutch residents into equally large deciles, or brackets, (D=1,2,\dots,10) such that the 10\% lowest-income households are grouped in the 1\ts{st} decile, and 10\% of the richest ones in the 10\ts{th} one.

\emph{Mixing matrix construction.} To operationalize the concept of the social opportunity structure, we introduce a mixing matrix, which is a matrix $X$ that represents the connectivity between different subgroups of the population as defined by their attributes. To construct such matrices, we count the occurrence of ties between households that belong to a certain subgroup as defined by one of the attributes in consideration and those in different subgroups of the population including their own. Based on these occurrences, we construct a mixing matrix in which rows and columns represent all possible values that an attribute takes. Each cell at the intersection of two subgroups of the population (for instance, two income deciles)  displays the share of contacts a household of a certain income bracket (as indicated by the value on the vertical axis) shares with the households in the income decile as indicated by the value on the horizontal axis. The elements of the matrix are normalized by row. The diagonal elements represent the share of intra-group contacts for each subgroup of the population. The matrices are therefore a visual representation of aggregated mixing patterns in the population that also serve as an input for further quantification of segregation levels.

\emph{Assortativity coefficient.} Finally, we quantify the level of segregation as assortative mixing along the chosen dimensions (income, highest achieved education level, ethnic group, and migration background). We do so through the common \emph{assortativity coefficient} $\rho$ \parencite{Newman2002, Newman2003} that captures the correlation between attributes of the connected nodes in a network. Based on previously obtained mixing matrices, the assortativity coefficient is defined as follows in Equation 1 for scalar or numeric values of income brackets, and in Equation 2 for discrete characteristics:

\begin{equation}\rho_{scalar}=\frac{\sum_{i, j} i j X_{i, j}-\sum_{i, j} i X_{i, j} \sum_{i, j} j X_{i, j}}{\sqrt{\sum_{i, j} i^2 X_{i, j}-\left(\sum_{i, j} i X_{i, j}\right)^2} \sqrt{\sum_{i, j} j^2 X_{i, j}-\left(\sum_{i, j} j X_{i, j}\right)^2}},\end{equation}

where \emph{i} and \emph{j} denote income deciles (D=1,2,\dots,10), \emph{X} represents previously defined mixing matrix normalized such that all elements add up to 1 and 
each element in the matrix \emph{X} represents a share of links between income decile \emph{i} and income decile \emph{j}; denominator presents the multiplication of the standard deviations of row-wise and column-wise sums;

\begin{equation}\rho_{discrete}=\frac{\sum_i X_{i i}-\sum_i a_i b_i}{1-\sum_i a_i b_i}=\frac{\operatorname{Tr} X-\left\|X^2\right\|}{1-\left\|X^2\right\|},\end{equation}

and \emph{a\textsubscript{i}}, \emph{b\textsubscript{i}} denote the row-wise and column-wise sum of normalized shares, respectively.

The assortativity coefficient $\rho$ essentially captures how diagonal a matrix is: higher observed assortativity would indicate a higher share of links within one’s own or similar subgroup of the population. The coefficient ranges from $-1$ to $1$, where higher positive values indicate a stronger tendency for mixing with someone of a similar trait. An assortativity coefficient of 0 signals random mixing, and negative values point to a tendency for matching with someone of a different type. While there are two assortativity coefficients defined for scalar and discrete values, for readability purposes, we refer to both as $\rho$. To contrast spatial and social network perspectives on segregation, we apply the measurement of the assortativity coefficient to both contexts. For the social network angle, we measure the assortativity coefficient of the social opportunity structure, i.e. direct network neighbors in a multilayer population-scale social network, referred to as “link assortativity”. For spatial analysis, we calculate the assortativity coefficient among neighbors as defined by the borders of the administrative neighborhoods (“buurt”).

The operationalization of segregation through nominal assortativity, however, has some limitations. First of all, nominal assortativity for both scalar and discrete values is influenced by the distribution of group sizes within the population of interest and possibly overlooks that smaller minority groups have relatively limited possibilities for creating stronger intra-group connectivity \parencite{karimi2023inadequacy, peel2020}. Second, degree distribution as well as network topology have strong implications for the observed levels of nominal assortativity. This is especially relevant for the scale-free networks in which some of the nodes have extremely high connectivity as compared to the rest of the network \parencite{Dorogovtsev_2010, karimi2023inadequacy, peel2020}. Finally, the way the measure is constructed in its classical definition does not account for the possibility of asymmetric mixing between different groups within the observed population \parencite{karimi2023inadequacy}. Altogether, the distribution of attributes within the population as well as network topology create additional constraints on the theoretical limits of the assortativity value. In an empirical scenario, the range of possible assortativity values is much narrower as compared to the theoretical limits from -1 to 1 \parencite{cohen, Cinelli2019}. 

Furthermore, it is necessary to highlight that the discrete and scalar assortativity values should be considered separately due to the differences in the nature of the analyzed features. For scalar values (income in this study) assortativity is calculated as the correlation between attribute values of the connected nodes, while the discrete version of assortativity (education level, migration, and ethnic background) is calculated based on modularity. What it implies is that for the calculation of a discrete version of assortativity we count only those occurrences where the two connected nodes have identical values. In the scalar version of assortativity, through the correlation formula, the relative differences between the values of the connected nodes are considered. Due to the nature of these variables in consideration, the differences between income groups can be objectively measured and quantified, which is not the case for any other features in the analysis (migration and ethnic background, as well as combined educational background). These considerations should be taken into account when interpreting the results as well as comparing nominal assortativity across different contexts. Nevertheless, despite these limitations, nominal assortativity remains a standard operationalization for homophily patterns and allows us to place our findings in a wider context of segregation literature.

To further shed light on problematic cases of higher concentration of contacts within one’s own subgroup, we introduce an additional metric that applies to scalar variables only (income deciles). We demonstrate \emph{within-decile concentration of contacts} in their own bracket as well as the two most closely related income groups. For instance, for households in decile 2 (second poorest income group), we represent the share of their contacts in the same, 2\ts{nd} income decile, and the share of their contacts that fall into the neighboring income decile (1\ts{st} and 3\ts{rd}). The expected value assuming uniform distribution among deciles would be that only 10\% and 30\% of one’s contacts would fall into their own and combined with two other most similar income groups, respectively. In the next section, we visually present the assortativity analysis in mixing matrices that show to what extent particular subgroups (based on their attributes) have an opportunity to interact with people who belong to the same or different subgroups. 

\section{Results}
In this section, we present the findings of our analysis, starting off by applying a segregation measurement strategy defined above to the selected node attributes that characterize households in the social network. We then focus on the most pronounced dimension of socio-economic segregation in social networks and contrast it with that of spatial neighborhoods. As a concluding step, we present assortativity analysis for the constituting layers of the social network.

\emph{Dimensions of segregation in the Netherlands.} First, we investigate segregation in socio-economic status or migrant background, arguably the two most studied facets of segregation. We capture segregation along the four dimensions: household income, highest achieved level of education, ethnic group, and migrant generation. 

\begin{figure}[h]
\centering
\includegraphics[width=\textwidth,scale=0.12]{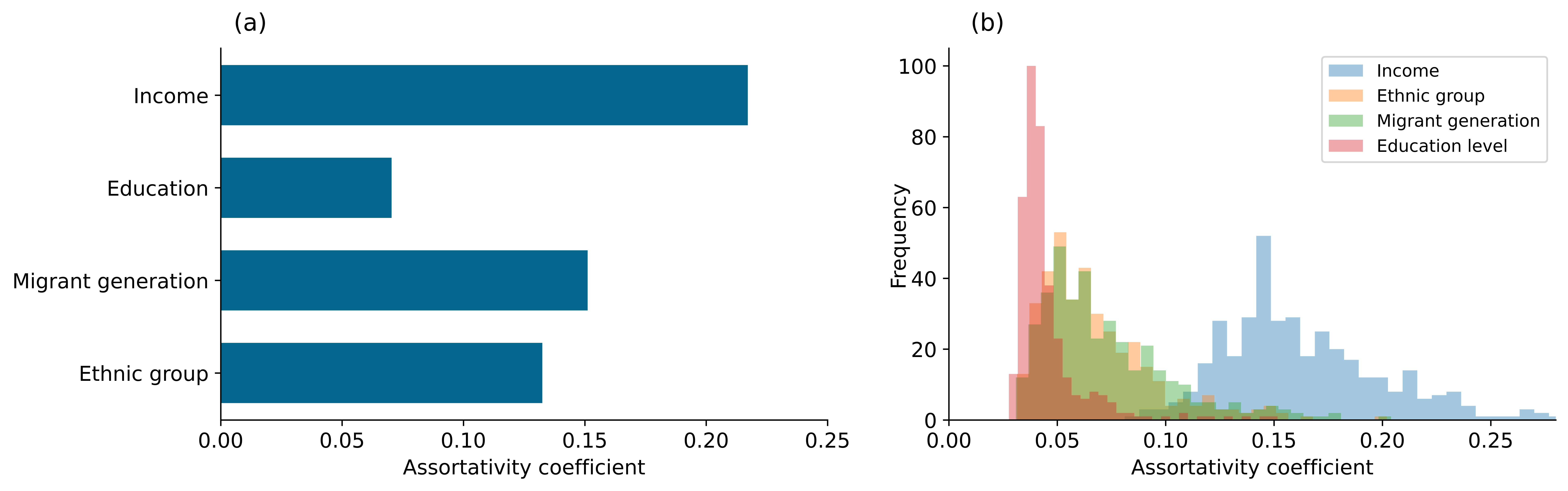}
\caption{\textbf{Assortativity in a population-scale network of households.} (a) Population-scale link assortativity values for various attributes. (b) Distribution of link assortativity across municipalities in the Netherlands: education level assortativity in red, ethnic group in orange, migrant generation in green, and income assortativity distribution is presented in blue.}
\label{fig:distributions}
\end{figure}

Figure~\ref{fig:distributions}a presents assortativity analysis of the multilayer population-scale social network that suggests that the captured social opportunity structure is segregated with respect to all four attributes, in the range of assortativity coefficient from around 0.05 to around 0.25. As an illustration, comparable previous work that uses assortativity to gauge socio-economic segregation finds values of no higher than around 0.5 even in the most extreme cases \parencite{Leo2016, Morales2019, Bokanyi2021, Dong2020}. For the scalar values, we find moderate levels of segregation along the income dimension (link assortativity $\rho$ = 0.22). On the other hand, we find that for discrete features, the highest achieved level of education is the least impactful trait in terms of segregating households, with an assortativity coefficient of only  0.07. This speaks to the importance of education as a means to break free from the household's own subgroup as defined by the educational background. The migrant background is again a clear segregator with an assortativity coefficient of 0.15 followed by 0.13 assortativity for the ethnic group attribute. 

The observed averages hide the considerable variation across different regions in the country. We therefore perform this assortativity analysis on the municipality-level social networks for each attribute. Figure~\ref{fig:distributions}b illustrates a great heterogeneity across the country with the distribution of assortativity values across all the municipalities in the Netherlands. The distribution of income assortativity by municipalities in the Netherlands in Figure~\ref{fig:distributions}b (the blue bars)  shows a wide range of values from 0.08 to 0.28. But also here income assortativity is consistently high for the majority of the municipalities.  Thus, despite a well-developed welfare state with a strong focus on income redistribution policies, income remains a significant cleavage in Dutch society. In what follows we therefore focus on income as the main dimension across which segregation manifests.

\begin{figure}[h]
\centering
\includegraphics[width=0.9\textwidth,scale=0.06]{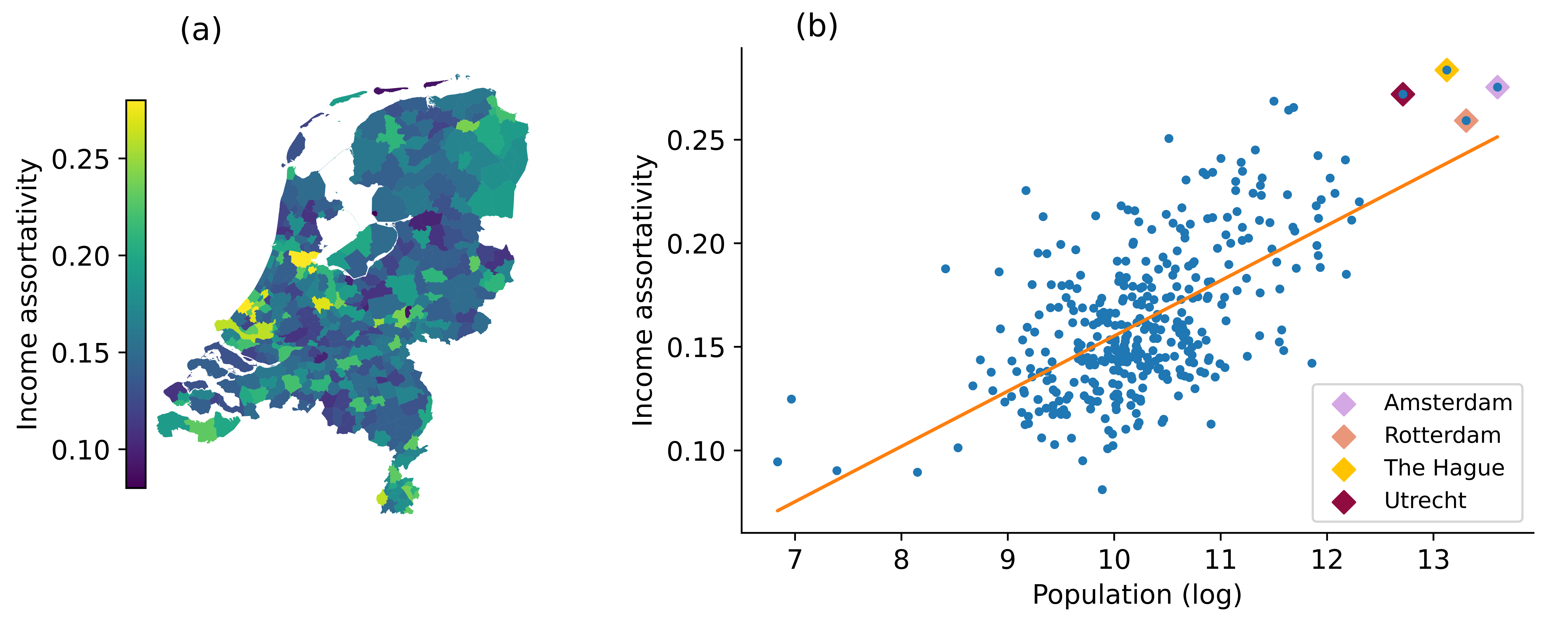}
\caption{\textbf{Geographical decomposition of income assortativity.} (a) Income assortativity per municipality. (b) Relationship between income assortativity and size of the municipality (natural logarithmic scale)}
\label{fig:geo_distribution}
\end{figure}

Figure~\ref{fig:geo_distribution}a shows how income assortativity values are geographically distributed across the country. Municipalities with the highest levels of income assortativity are located mostly in the central-western Netherlands, the so-called “Randstad” area containing the largest Dutch cities including Amsterdam, The Hague, Rotterdam, and Utrecht, and their suburbs. Similarly, we see that the border regions of Zeeland (bottom left) and of the South of Limburg (bottom right) that contain some of the larger municipalities (Maastricht, Terneuzen) also exhibit relatively high levels of income assortativity. In fact, we find that there is a log-linear relationship between income assortativity in a municipality and the size of its population (Figure  ~\ref{fig:geo_distribution}b). This means that contrary to smart urban planning approaches perception of cities as “champions of diversity” we find that larger municipalities exhibit higher levels of segregation. This is not due to differences in degree distributions across various municipalities as they are consistent, no matter the population size (see Supplementary Figure ~\ref{fig:hh_distributions_combined}).

\begin{figure}[h]
\centering
\includegraphics[width=0.9\textwidth,scale=0.06]{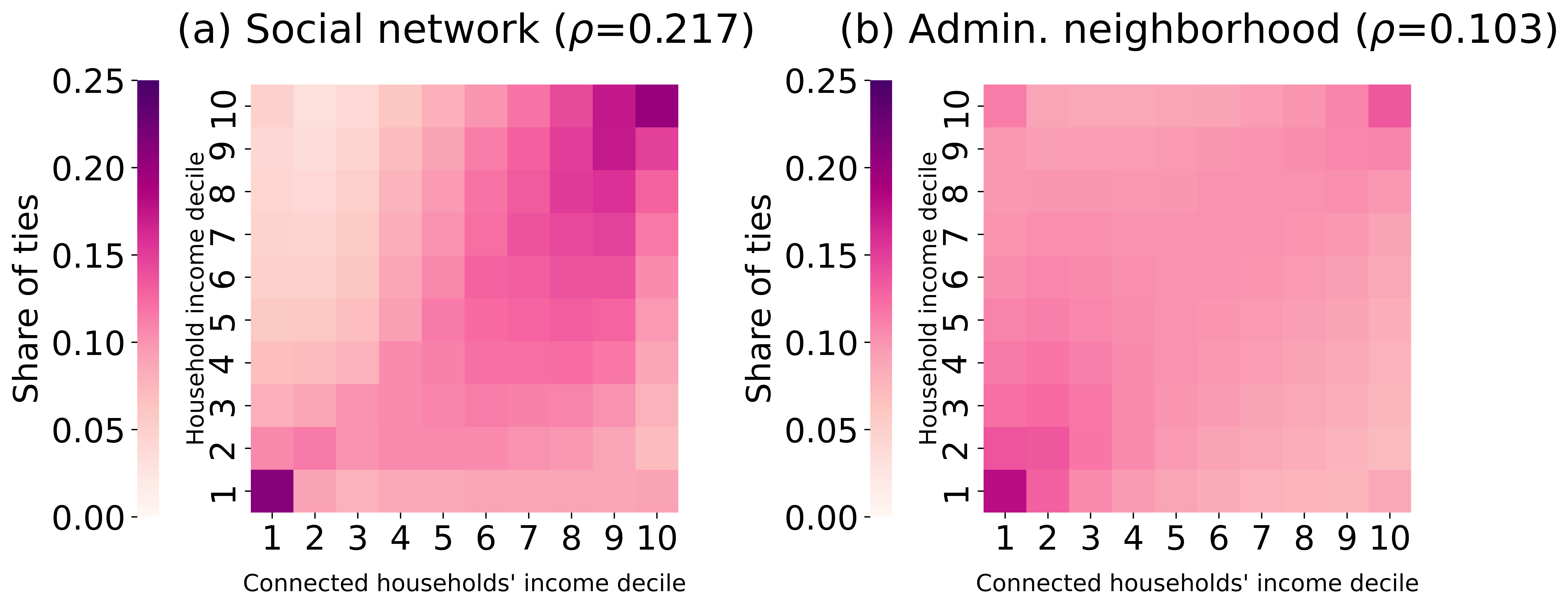}
\caption{\textbf{Mixing matrices} for population-scale social network of households (a) and administrative neighborhoods (b). Colors denote the share of links the two income deciles share.}
\label{fig:main_comparison}
\end{figure}

\emph{Comparing the social network and the spatial approach.} We now turn to the main issue at hand: how segregation in the social network compares to segregation in spatial neighborhoods. Figure~\ref{fig:main_comparison} shows the social opportunity structures for the population-scale social network and the administrative neighborhood approaches, as introduced and discussed in Section \ref{sec:methods}. The comparison of the two reveals striking differences. First and foremost, segregation in the social network is more than twice as much as in administrative  (spatial) neighborhoods (income assortativity resp. 0.217 and 0.103). Second, the distributions of connections across income ranges are markedly different. The social network perspective (Figure~\ref{fig:main_comparison}a) displays a higher concentration of contacts across the diagonal of the matrix and neighboring cells. This indicates a stronger tendency of households to share ties with other households that are (relatively) similar in terms of income. On the contrary, the administrative neighborhood matrix (Figure~\ref{fig:main_comparison}b) shows a rather uniform distribution of contacts across the entire income range. The only exception is the relatively high concentration of contacts among the poorest as well as amongst the richest deciles. Yet, even this self-orientation at the extremes is still more pronounced in the social network matrix (by 3 percentage points for the bottom-left cell for the poorest decile and 6.5 percentage points for the top-right cell for the richest one). Furthermore, administrative neighborhoods demonstrate slightly higher exposure between the extremes of income distribution. The top-income decile has a relatively higher share of contacts with the poorest-income decile which is partially driven by policy efforts aimed at evenly redistributing social housing units across the city, even in the affluent neighborhoods.

Population-scale social network patterns of segregation are clearly more pronounced compared to spatial patterns of segregation. While households may well have a rather diverse exposure to households of various socio-economic backgrounds in spatial neighborhoods, their actual social network opportunity structure shows consistently higher levels of segregation. In fact, people are more than twice as likely to be socially confined in their own income group in the network as compared to spatial neighborhoods. Furthermore, if we consider segregation through the spatial neighborhood approach, the finding suggests that segregation occurs predominantly at the extremes of the income distribution --- most evidently for the poorest and the richest income deciles. Nonetheless, in addition to the high levels of income segregation for the poorest and the richest income groups, social networks reveal moderate levels of segregation, but across the whole income range. 

\begin{figure}[h]
\centering
\includegraphics[width=\textwidth]{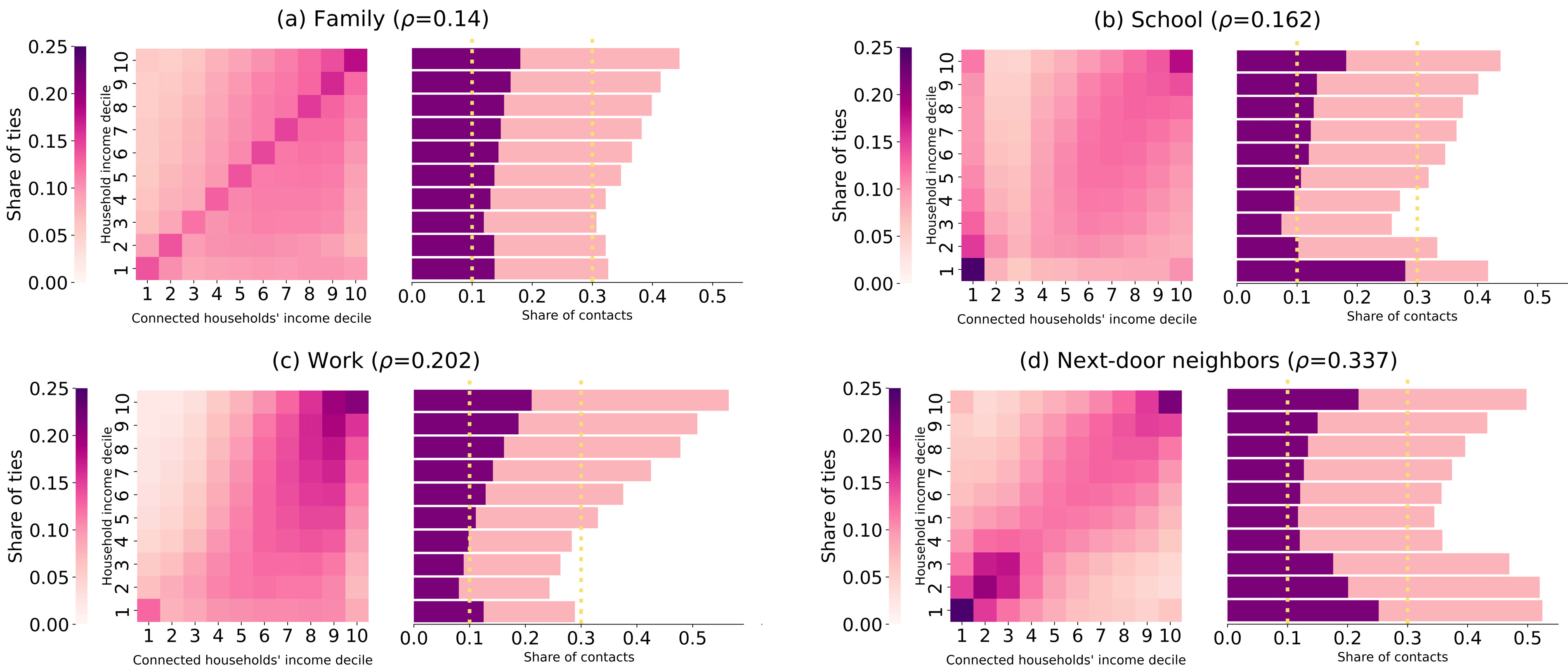}

\caption{\textbf{Assortativity for four layers:} (a) family, (b) school, (c) work, (d) next-door neighbors. The left-hand side of each panel contains the mixing matrices, similar to Figure~\ref{fig:main_comparison}. The right-hand side of each panel presents the within-decile concentration of contacts within similar income groups: dark purple bars represent the concentration of contacts within the exact same income decile, and pink bars show the concentration of contacts within the two other most similar income deciles. Yellow dashed vertical lines at 0.1 and 0.3.
}
\label{fig:layers_comparison}
\end{figure}

\emph{The multi-layered nature of social segregation.} We have established that social segregation is more pronounced if we study it through a lens of social opportunity structure. This leads to the question of how the constituent layers or social contexts of that social opportunity structure reinforce social segregation. We, therefore, calculate the contribution of each layer to the overall income assortativity pattern observed across the whole network. We distinguish between the layers discussed in Section \ref{sec:methods}, namely: family, school, work, and next-door neighbors, and find rather dissimilar segregation patterns, as Figure~\ref{fig:layers_comparison} illustrates. 

\begin{figure}[h]
\centering
\includegraphics[width=\textwidth,scale=0.02]{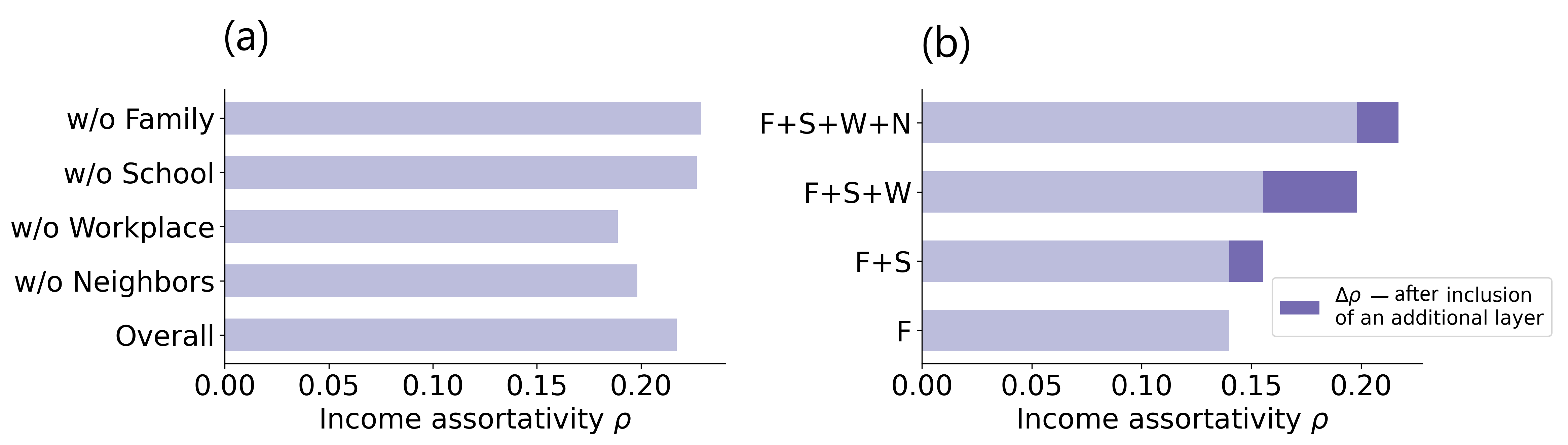}
\caption{\textbf{Contribution of each layer to the overall level of income assortativity:} \\ (a) income assortativity levels for network composed by alternating exclusion of individual layers (w/o (without) denotes the layer which was excluded from the network), (b) income assortativity levels for network composed by consecutive inclusion of individual layers.}
\label{fig:robust_combined}
\end{figure}

Family is the least socio-economically segregated context ($\rho$=0.14), followed by school ($\rho$=0.16), work ($\rho$=0.20), and finally next-door neighbors, the most segregated network layer ($\rho$=0.34). This ranking is consistent across the whole country, irrespective of the population size of the municipality (See Supplementary Figure ~\ref{fig:graph_layers}). When combined, these layers constitute the segregated social opportunity structure as presented in Figure~\ref{fig:layers_comparison} ($\rho$=0.22). When we remove one of the layers from the entire social opportunity structure, the segregation score remains in the range of 0.19 to 0.24, which indicates that there is not one layer that predominantly drives segregation (Supplementary Figure ~\ref{fig:robust_combined}a). When we measure segregation values by adding layers on top of each other in the order mentioned above, we notice that the work layer makes the largest contribution to segregation (Supplementary Figure ~\ref{fig:robust_combined}b). This is not unexpected given that the work layer is also the largest one in terms of average degree.

Figure~\ref{fig:layers_comparison} presents a detailed assortativity analysis for each of the four layers of the network. For each matrix, the right-hand side bars show the exposure to households in one’s own income decile (purple bar) for each decile as well as additional exposure to households in the two other most closely related income groups (pink bars). The yellow vertical dashed lines give a 10\% and 30\% expected threshold under the uniform distribution of ties over the entire population. This reveals additional insights into the pattern of segregation for each layer. Family connections (Figure~\ref{fig:layers_comparison}a, left), for instance, are the least segregated in overall terms as indicated by the assortativity coefficient of 0.14. Still, they have an apparent diagonal that indicates that across the whole income range, the highest number of family ties is with households of the exact same socio-economic standing. This is well illustrated by the right-hand side of Figure~\ref{fig:layers_comparison}a, which shows that across the whole income range, exposure to one’s own income decile is higher than the 10\% baseline. When we also take into account the most similar income groups represented by pink bars (for instance, for the poorest income decile 0, the two most similar income deciles beyond its own are decile 1 and 2; for decile 3, these are 2 and 4), it is clear that the richer the household is, the more likely it is to have family members of either exact same or similar income background. The top income decile even has half of its social exposure through family connections to the three most similar income groups.

The school layer exhibits a slightly higher level of segregation but a highly dissimilar pattern compared to the family layer (Figure~\ref{fig:layers_comparison}b - left). In the school context, all households have relatively high exposure (on average, 15\% of all contacts) to the bottom 10\% poorest households. This exposure is the highest for the lowest income decile. About 30\% of households connected to the lowest-income households are equally low-income households. In a similar way, the top 10\% richest households in the country have the highest concentration of school connections within their own income decile. Moving to the most similar income deciles, segregation becomes even more pronounced, with the top and bottom deciles having around 50\% and around 45\% of their social opportunities within their own income range.  The granularity of our data allows us to further specify segregation patterns by school type. This relatively high level of segregation at the extremes of the income distribution (poor to poor as well as rich to rich) is reproduced in primary and secondary schools, whereas vocational training institutions create high exposure primarily among the poorest students. Unlike the other institutions, higher education exhibits high exposure to the bottom 10\% poorest households across the entire income range (corresponding matrices are presented in Supplementary Figure ~\ref{fig:school_layers_combined}). This is due to the fact that in many cases children in a household move out and become independent single-person households when they join a higher education institution. And these single-person households to a large extent are without income and fall into the poorest income decile in the population.

While the workplace social opportunity structure (Figure~\ref{fig:layers_comparison}c) does not exhibit an apparent diagonal of the matrix as in, for instance, the family layer, there is a general tendency of having work colleagues if not in the exact same then in a relatively similar income bracket. Moreover, this pattern is stronger for higher income range households. The upper half of the income distribution has a disproportionately high exposure to households of similar socio-economic standing, and this tendency becomes more pronounced with an increase in household income.

The most segregated network layer is that of next-door neighbors (Figure~\ref{fig:layers_comparison}d), and this is mostly due to the bottom 30\% of the income range as well as the top 10\% richest households with around 20-25\% of next-door neighbors being in the exact same income decile as the observed households. For the middle as well as upper-middle income classes, the tendency for mixing within households’ income decile or similar income groups is present, yet less pronounced. This may well reflect the opportunities for the higher income deciles to choose where they want to live, as well as the inability of the lowest incomes to do so. 

\begin{figure}[h]
\centering
\includegraphics[width=0.6\textwidth,scale=0.25]{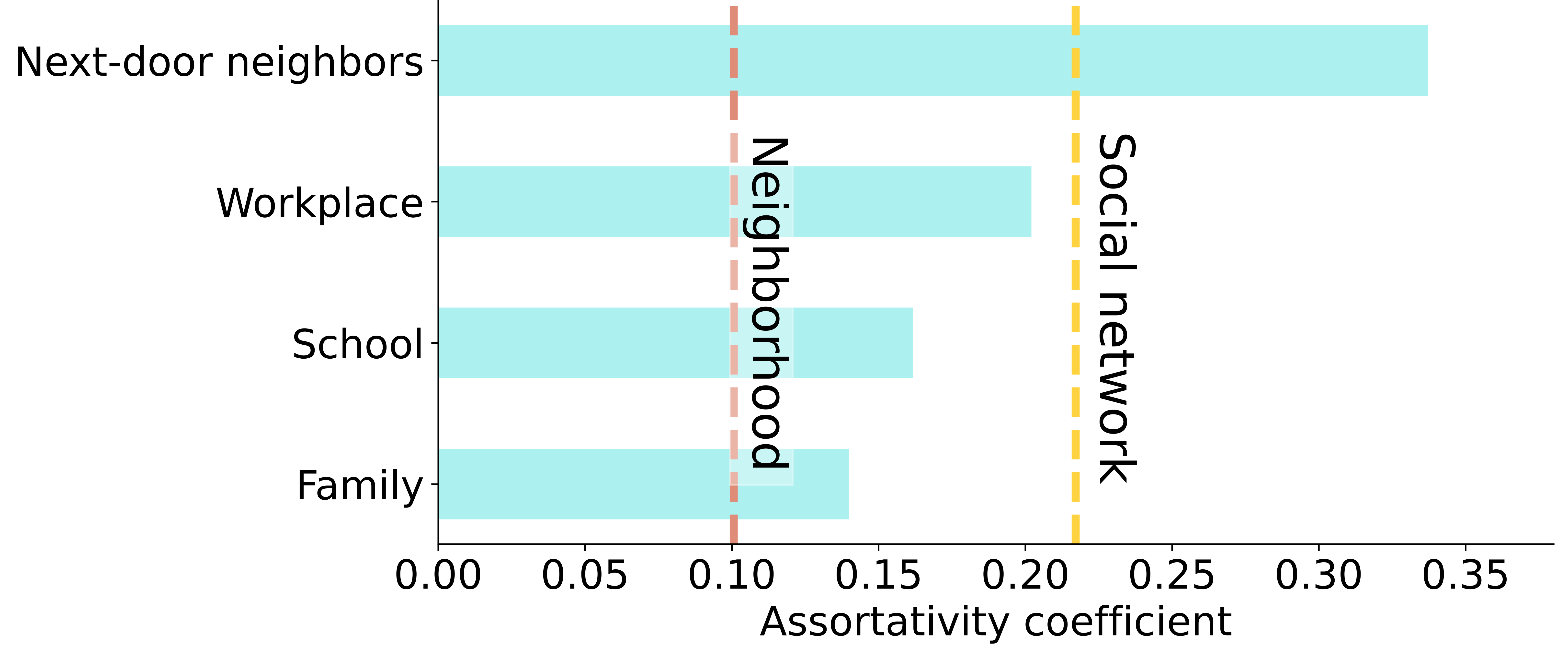}
\caption{\textbf{Income assortativity} for administrative (spatial) neighborhoods (orange dotted line), population-scale social network (yellow dotted line), for each layer (next-door neighbors, workplaces, school, and family).}
\label{fig:concluding_figure}
\end{figure}

Figure~\ref{fig:concluding_figure} visually summarizes the results of our analysis and shows the level of segregation as measured by assortativity for each layer, for the network as a whole, and for segregation measured through administrative neighborhoods. The population-scale social network approach introduced here exhibits a segregation twice as high as could be captured across the boundaries of administrative (spatial) neighborhoods. This is apparent in each constituting network layer --- family, school classmates, work colleagues, and next-door neighbors. This leads to the conclusion that at the level of the traditional unit of spatial aggregation (e.g. administrative neighborhoods), patterns of socio-economic segregation are minimized, while in fact, they persist in the underlying structure of social networks.

%

\section{Discussion and conclusion}

We asked the question of whether patterns of segregation are more pronounced in social networks than the common spatial manifestations of segregation. Our research showed that levels of segregation in the population-scale social network are indeed more than twice as high as in administrative neighborhoods. Neighborhood-level indicators of segregation hide underlying persistent patterns of segregation and inequalities. This means that both scholars and policymakers may systematically underestimate the levels of segregation in society. Of course, spatial and social aspects remain deeply intertwined and have to be considered complementarily.  Our contribution is to point out that neighborhoods may capture some but certainly not all relevant parts of social ties that segregate our societies. As such, we position ourselves in the literature that calls scholars and policymakers to go beyond spatial patterns of segregation. We suggest a practical and low-cost approach to include social networks, while others have, for instance, convincingly argued to include temporal segregation (i.e. how segregation varies throughout a day, a week, and different seasons \parencite{Silm2014}), segregation in amenities \parencite{Moro2021}, or social interactions \parencite{Calvano2022}, as well as segregation measurement based multi-hop distances in the network \parencite{laan_emery_2021}.  Going beyond spatial patterns is necessary to understand the persistent inequalities in our societies.

Our approach also allows us to further investigate segregation as a multi-layered phenomenon that plays out across different sets of social contexts. We found that social contexts are rather dissimilar in terms of socio-economic segregation. The most diverse exposure is presented by family and school layers. Family connections are the least segregated social context, yet we still observe a higher concentration of contacts within one's own income group across the whole income range. And this pattern is more pronounced for affluent households. The school layer, marginally more segregated as compared to family, reveals socio-economic segregation at the extremes of the income distribution. This is to a large extent a result of segregation patterns in lower education levels, primary and secondary schools. Previous research has provided evidence that for earlier stages of education trajectories, especially elementary school, due to relatively low commuting distances the school segregation pattern is to a large extent a reflection of residential segregation \parencite{Boterman2019}. On the other hand, at later stages of education, students have a higher likelihood of commuting longer distances to a school of choice or even moving out of a parent’s household to pursue further education. This is reflected by a high exposure to poor households in the context of higher education: children that moved out of their parents’ house for the purposes of studying become independent, single-person households with low to no income, thus we observe the highest exposure to this kind of low-income students across the whole income range. It has been also pointed out by \cite{laan_emery_2021} that segregation by education is heterogeneous across groups of people of different education levels: holders of a bachelor's degree or its equivalent is the least segregated subgroup, while people with a master's degree or higher are the most isolated subset of the population \parencite{cbs_basdoard}.

On the other hand, work colleagues as well as next-door neighbors are the social contexts that limit one’s exposure to diverse socio-economic groups most. Colleague ties represent the most numerous (in terms of average degree), but also the second most segregated social context. Social mixing patterns at workplaces demonstrate limited exposure to various socio-economic groups and this is the most prominent for the richer half of the income distribution. Finally, the most segregated network layer among all is next-door neighbors, with the extremes of the income distribution --- the poorest and the richest households --- being the most segregated. This speaks to the low effectiveness of city planning practices aimed at creating more diverse neighborhoods. Even though the composition of the neighborhood as defined by administrative boundaries may seem well-mixed, on a more granular level, people living next door are very likely to belong to one’s exact same socio-economic strata. Such a difference in segregation level of administrative neighborhoods vs next-door neighbors arises given the property/rental price in a specific housing unit (in the case of heavily urbanized areas with high-density housing), while the property profile of the whole neighborhood is more diverse and can include a wide variety of properties of different price/rent ranges as well as social housing units \parencite{Reardon2008}. This is further reinforced by a common policy objective in some municipalities in the Netherlands to redistribute social housing units evenly which leads to the situation when even the most expensive neighborhoods within their respective administrative boundaries have some of the social housing units allocated \parencite{HousingSupplyIMF}. 

The aim of our study was to compare two different approaches to capturing socio-economic segregation and not to establish whether segregation in the Netherlands is high or low. Our findings can of course be placed in the context of other work. The overall level of income assortativity of 0.22 we found ties in well with previous studies wherein empirical social networks exhibit similar assortativity levels. A recent study on patterns of urban mobility and social network structure based on Twitter data from the fifty largest municipalities in the USA report distributions of income assortativity in the range from 0.05 to 0.5, with the majority of metropolitan areas falling into a narrower 0.1--0.25 span \parencite{Bokanyi2021}. Other work on empirical interaction networks (inferred from purchase and Twitter data) at a country level yields up to around 0.5 income assortativity \parencite{Dong2020}. The levels of segregation we found seem very realistic and a good foundation for further comparative work, both across countries and over time. Of course, such comparative studies need to carefully consider the sensitivity of the assortativity measure to the network structure as well as the underlying inequality in income distribution.  

The network that we investigated here is a social opportunity structure. We can therefore assume that social networks where tie formation or activation is based on user choice are even more segregated. The interaction between the social opportunity network and activated ties is therefore a fruitful and important avenue for additional research. Of particular interest here is enriching population-scale social network data from registers with population-wide surveys that include questions on the structure and composition of social networks. 

Our findings also speak to the perception of cities as “champions of diversity”. We find that larger municipalities tend to exhibit higher levels of socio-economic segregation.  In larger cities, despite a diverse pool of social contact opportunities, households gravitate towards their own income range in contexts such as family, school, workplace, and next-door neighbors.  An intuitive reason for this finding is that in smaller communities there is a relatively narrower pool of choices in terms of affiliations and individuals are hence limited in choosing the kind of social ties they desire (low choice homophily). This moderates segregation in smaller communities. On the contrary, larger cities, despite a more diverse demographic distribution, have a greater potential for choice homophily to materialize. A similar pattern of results was obtained in a CBS' project on educational exposure which uncovered that very strongly urbanized environments exhibit the highest levels of segregation for certain subgroups of the population, namely those with lower education and with a master's degree or higher \parencite{laan_emery_2021, cbs_basdoard}. Comparably, a recent study based on a social interaction network inferred from mobile communication data \parencite{Nilforoshan2022} demonstrated that in terms of potential for social interactions larger metropolitan areas are more segregated. Authors theorize the observed pattern is a result of a more diverse socio-economic differentiation of amenities. Even though our study relies upon a vastly different approach to social network inference, these results complement each other and both challenge a dominant view of cities as hubs for diverse socio-economic mixing.

For the case of the Netherlands, we established that segregation is still pronounced along the dimension of household income. Nevertheless, in the public and policy discourse on segregation, it often fades against the background of “ethnic” segregation, i.e. segregation of native vs multiple migrant categories of population \parencite{Boterman2021}. Interestingly, in the Dutch context, income-based interventions are typically used as a proxy for tackling segregation associated with immigration. This is because ethnic segregation raises various ethical and political concerns, and also because neighborhood income is associated with its ethnic composition \parencite{vangent2016}.  However, we found that this correlation between ethnic background and income does not hold at the household level (as reported in Supplementary Table ~\ref{tab:attr_correlation}). This leads to the somewhat surprising situation that while the policy interventions to reduce “ethnic” segregation are ill-informed and therefore misdirected, these same interventions do in fact by accident impact the more important dimension of socio-economic segregation.  

Our approach may inspire new avenues for effective policies to combat segregation and related inequalities. Such avenues should look beyond the assortativity values that we report for the network and its constituting layers and at least take into account the number of connections in each layer as well. For instance, effective policies that decrease workplace segregation may bring more benefits, since it affects more connections than decreasing, for instance, next-door neighbor segregation, even if income assortativity values for the latter are higher. 

It is also worth exploring the gap between low segregation levels observed within the boundaries of administrative neighborhoods and the three times higher segregation level among next-door neighbors. We demonstrated that neighborhoods (within their administrative boundaries) provide diverse exposure to a variety of socio-economic groups, however, there is no direct evidence that would support the translation of such exposure into social contact with these different socio-economic groups. So, on one hand, observed low levels of segregation in administrative neighborhoods can be regarded as the success of housing policy as they create diverse exposure, however, the diversity of exposure does not guarantee the diversity of the social network. Exploring the relationship between the increased diversity in neighborhoods and changes in the social network composition associated with it presents a promising research direction that would, however, require a longitudinal analysis of the changes in neighborhood and social network compositions simultaneously. 

The analysis conducted in this study also creates foundations for conceptualizing and measuring social capital in the context of a population-scale social network. Taking into account a wider social opportunity structure as opposed to a more traditional network of close contacts obtained through surveys, or name/position generators potentially allows for a more comprehensive and broader evaluation of “resources” embedded in social structures as well as a possibility to capture structural aspects of these social networks. Furthermore, additional data collection based on similar registers in other countries would make comparative work across borders possible. Leveraging register data from other contexts and geographies would allow for an international comparison of segregation levels given different social policies and welfare regimes and help evaluate the efficiency of policy tools aimed at reducing segregation.

The approach used by us also comes with a number of limitations. First of all, although the data coverage and quality are unprecedented, they are not flawless. The family layer for instance is impacted by missing family registers before 1995 and a large share of missing family ties for first-generation migrants. The work layer currently suffers from a set boundary on the firm size that applies to large companies. While most of the companies in the country are up to 100 people (the overwhelming majority of the employers (96.26\%) have less or equal to 50 employees; 98.04\% fall below 100 people personnel), rare large corporations with several thousand employees do not currently provide a department-level organizational structure that could improve the sampling for this subset of employers. And the school layer cannot distinguish between school classes in the same year. These are not only important caveats to take into account when interpreting the results but also a necessary input for improving such population-scale social network datasets. 

Second, registers only capture “formal” ties and miss out on more “informal” ties such as friendship. However, previous work has firmly established that friendships are mostly realized from these contextual relationships to institutions. Nevertheless, we expect that informal ties such as friendships exhibit more choice homophily and add to the levels of segregation. Third, the population-scale social network is strictly defined by the borders of the country, thus it omits ties that span across countries. While we checked for consistency of degree distributions between different parts of the country, this is highly relevant, especially for border regions, where people could commute to a neighboring country for work or might have a family there, or for those cases when Dutch residents migrated elsewhere.

Finally, as we mentioned in Section \ref{sec:methods} the nominal assortativity measure has some limitations, namely it ignores the distribution of group sizes and is heavily biased by scale-free degree distribution and asymmetric mixing patterns \parencite{peel2020, karimi2023inadequacy, Cinelli2019, cohen}. In relation to this, we treat the main attribute of interest – the household income – by splitting the full income range of Dutch residents into equally large groups. So the calculations on income assortativity in the whole population-scale social network as well as in its individual layers are based on evenly sized groups of income that ensure adequacy of the used nominal assortativity measure. In addition to that, the population-scale social network in consideration is highly distinct from the random or scale-free structures that typically suffer from inadequacy of nominal assortativity. Unlike scale-free structures with the presence of a small number of nodes with a relatively high number of outgoing links (“stubs”), the degree distribution of the population-scale social network follows none of the most commonly assumed shapes, i.e. a power-law or a lognormal distribution \parencite{Bokanyi2022}. Per-layer degree as well as total degree distributions indicate that the overwhelming majority of the nodes have an average or close to an average degree, and the presence of high-degree nodes is extremely low. Based on this, we conclude that the risk of a bias in the observed level of assortativity due to skewness of degree distribution is minimized as compared to scale-free structures that are typically discussed in this context. 

Furthermore, there is a possibility of asymmetry in mixing patterns as well as a higher prevalence of relatively higher-degree nodes in certain subgroups of the population \parencite{karimi2023inadequacy}. For instance, it might be the case that, on average, native Dutch residents have a higher degree in the family layer, as compared to the migrant subgroup of the population, especially the first generation. While this is partially tackled by the sampling rules and thresholding which is applied universally to all individuals in consideration, future research could continue to explore the asymmetry in observed mixing patterns. Such extension would go beyond the current goal of capturing and estimating the level of homophily along multiple dimensions in the population-scale social network and pertain to the generative aspect of the observed segregation levels. 

Our results suggest that social networks play a fundamental role in assessing socio-economic segregation and uncovering its multi-layered nature. Focusing on population-scale social networks overall as well as its constituting layers reveals remarkably consistent higher levels of segregation as compared to what could be captured with existing approaches. This may be considered a promising validation of incorporating social-network-aware measures of segregation into existing social cohesion evaluation frameworks. Furthermore, such extension is becoming more achievable in light of the increasing availability of population registers worldwide.

\section{Ethical statement}

The register dataset used for creating the Dutch population-scale network is pseudonymized before research access, and it does not contain personal information such as address or date of birth. Still, because of the sensitive nature of the information, storage, and analysis were done within the secure server environments of Statistics Netherlands. Further details about data generation, ensuring data security and privacy are described in \cite{VanderLaan2021}.

{
\printbibliography
}

\vspace{180mm} 
\hfill \break
\section {Supplementary information}
\beginsupplement

\begin{figure}[h]
\centering
\includegraphics[height = 17cm]{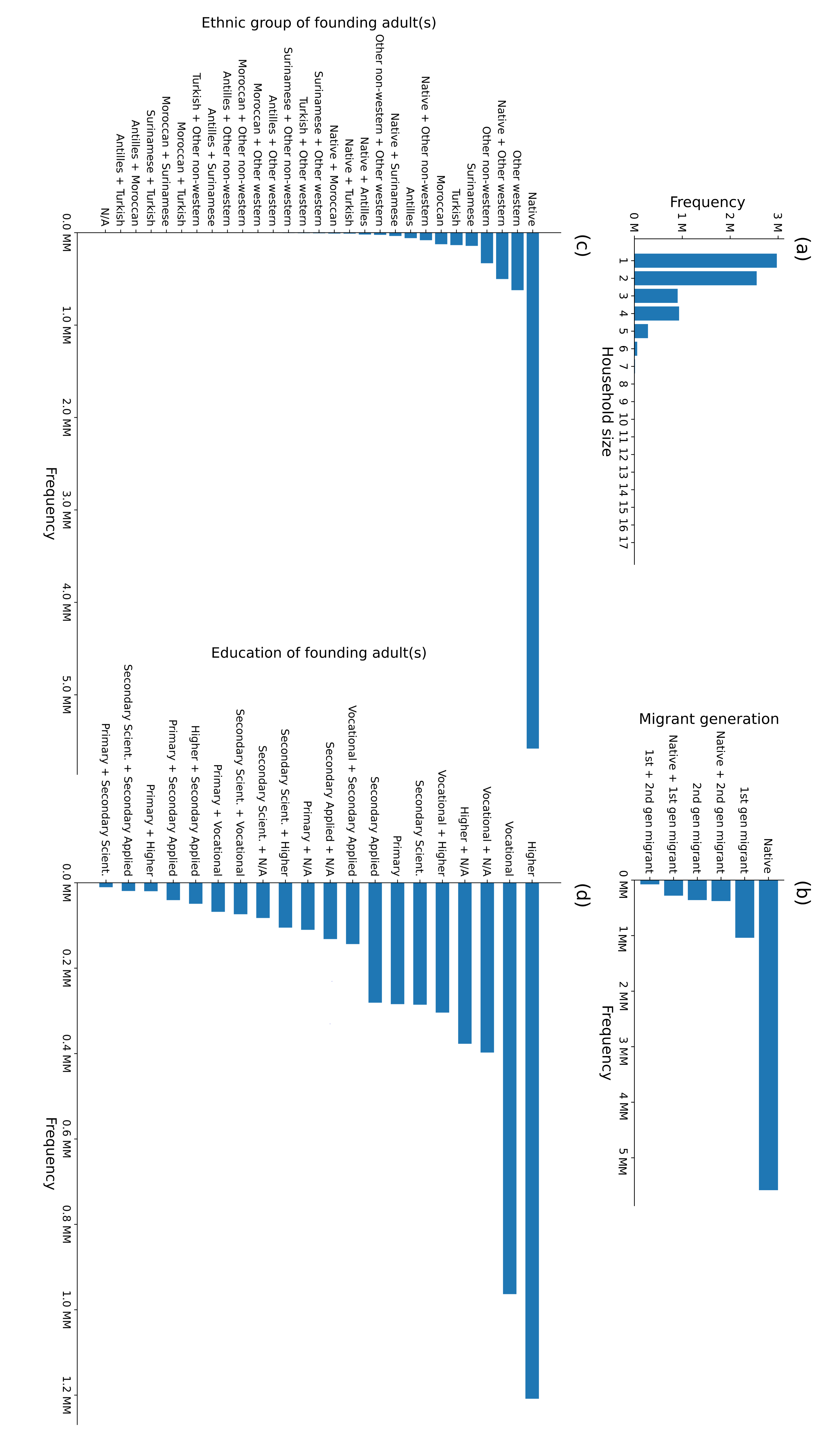}
\caption{\textbf{Distribution of household attributes:} (a) distribution of the households by size, (b) migration generation, (c) ethnic composition of the founding adult(s), and (d) their educational background.
}
\label{fig:hh_distributions_combined}
\end{figure}

\begin{figure}[h]
\centering
\includegraphics[width=0.7\textwidth,scale=0.005]{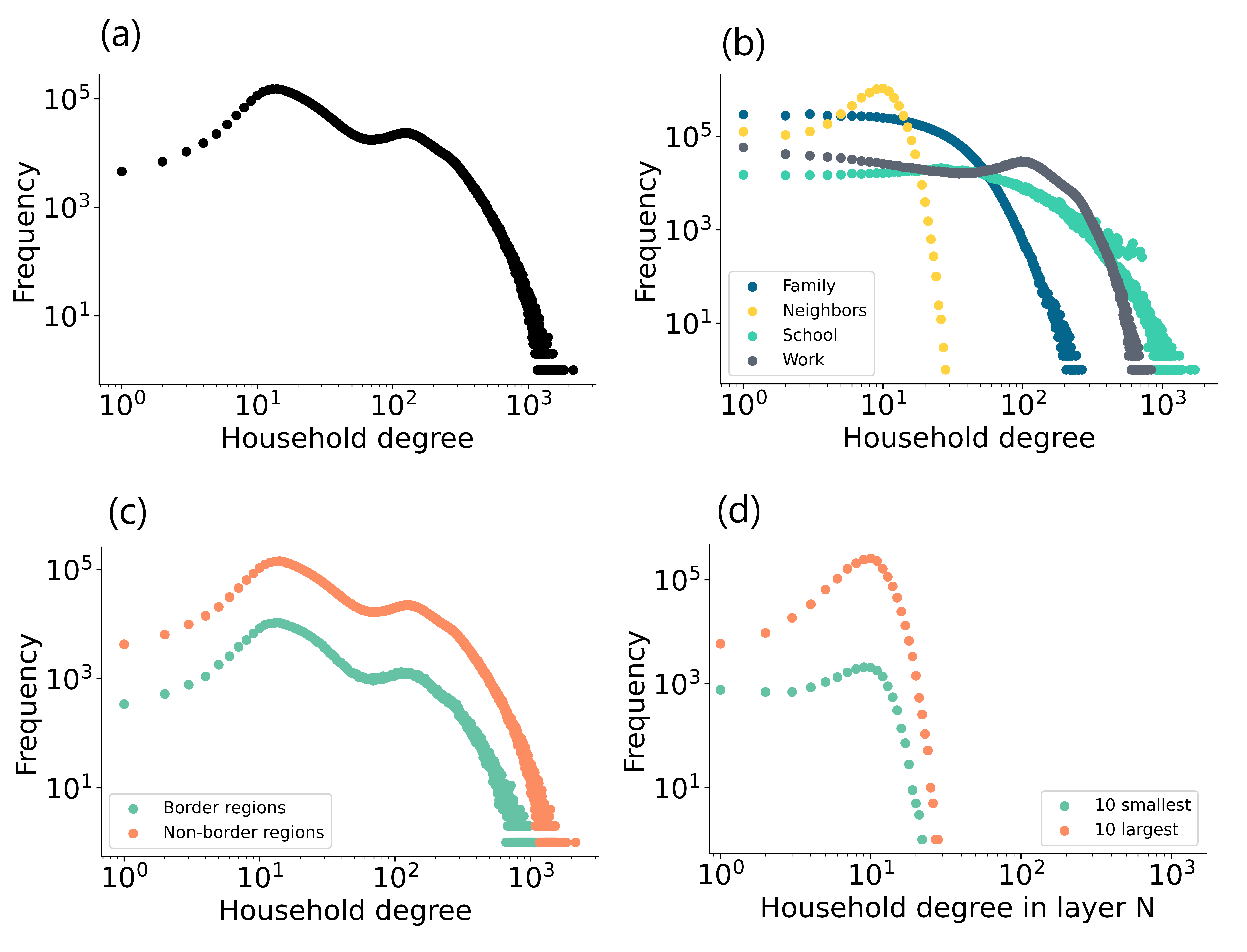}
\caption{\textbf{Household degree distributions.} (a) Overall household degree distribution. \\ (b) Household degree distribution for individual layers of the network: family, next-door neighbors, school, and work. (c) Overall household degree distribution for two groups of municipalities: border regions (green dots) and the rest of the country (orange dots). (d) Distributions of the household degree in the neighbors layer for the largest (orange dots) and the smallest municipalities in the country (green dots).}
\label{fig:degre3es_combined}
\end{figure}

\begin{table}[htb]
\small
\centering
\begin{tabular}{@{}|r|l|c|c|c|@{}}
\hline
\multicolumn{1}{|l|}{} &

  \multicolumn{1}{c|}{\emph{Income decile}} &
  \emph{Migrant generation} &
  \emph{Ethnic group} &
  \emph{Education level} \\ 
\hline
\emph{Income decile} &  \cellcolor[HTML]{C0C0C0}
   &
 \cellcolor[HTML]{FFCCC9} \begin{tabular}[c]{@{}c@{}}P-value for ANOVA\\ is 0.045,  accept H\textsubscript{0}:\\ variables are not \\ correlated.\end{tabular} &
\cellcolor[HTML]{FFCCC9}  \begin{tabular}[c]{@{}c@{}}P-value for ANOVA \\is 0.174, accept H\textsubscript{0}:\\ variables are not \\ correlated.\end{tabular} &
\cellcolor[HTML]{FFCCC9}  \begin{tabular}[c]{@{}c@{}}P-value for ANOVA\\ is 0.108, accept H\textsubscript{0}:\\  variables are not \\ correlated.\end{tabular} \\ 
\hline
\emph{Migrant generation} &  \cellcolor[HTML]{C0C0C0}
   &  \cellcolor[HTML]{C0C0C0}
   &
  \cellcolor[HTML]{C0C0C0} N/A &
 \cellcolor[HTML]{CDF5CD}  \begin{tabular}[c]{@{}c@{}}P-value for Chi-square \\test is 0.0:  variables\\ are correlated.\end{tabular} \\ 

\hline
\emph{Ethnic group} &  \cellcolor[HTML]{C0C0C0}
   &  \cellcolor[HTML]{C0C0C0}
   &  \cellcolor[HTML]{C0C0C0}
   &
  \cellcolor[HTML]{CDF5CD}  \begin{tabular}[c]{@{}c@{}}P-value for Chi-square\\ test is 0.0: variables\\ are  correlated.\end{tabular} \\ 
\hline
\emph{Education level} &  \cellcolor[HTML]{C0C0C0}
   &  \cellcolor[HTML]{C0C0C0}
   &  \cellcolor[HTML]{C0C0C0}
   &  \cellcolor[HTML]{C0C0C0}
   \\ 
\hline
\end{tabular}
\caption{{\textbf{Results of the significance tests for correlation between household attributes:} p-values for ANOVA to test the correlation between a numeric variable (income decile) and the remaining categorical variables (migrant generation, ethnic group, education level) as well as p-values of Chi-square test to examine the strength of the relationship between categorical variables. The confidence interval is 99\%.
}}
\label{tab:attr_correlation}
\end{table}

\begin{figure}[h]
\centering
\includegraphics[width=0.7\textwidth,scale=0.005]{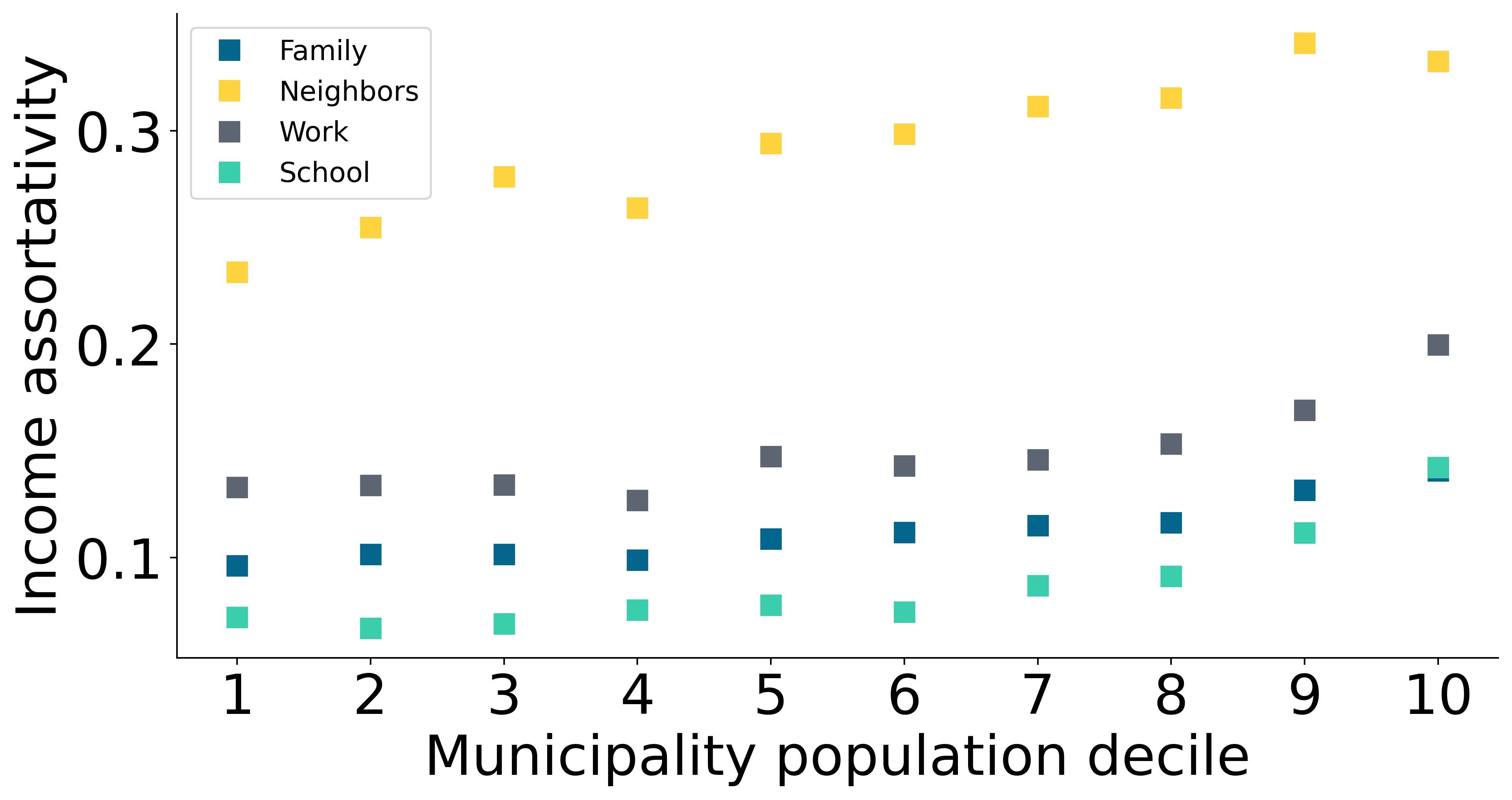}
\caption{\textbf{Scatterplot of income assortativity} per layer in relation to the population size of a municipality.
}
\label{fig:graph_layers}
\end{figure}

\begin{figure}[h]
\centering
\includegraphics[width=0.9\textwidth,scale=0.02]{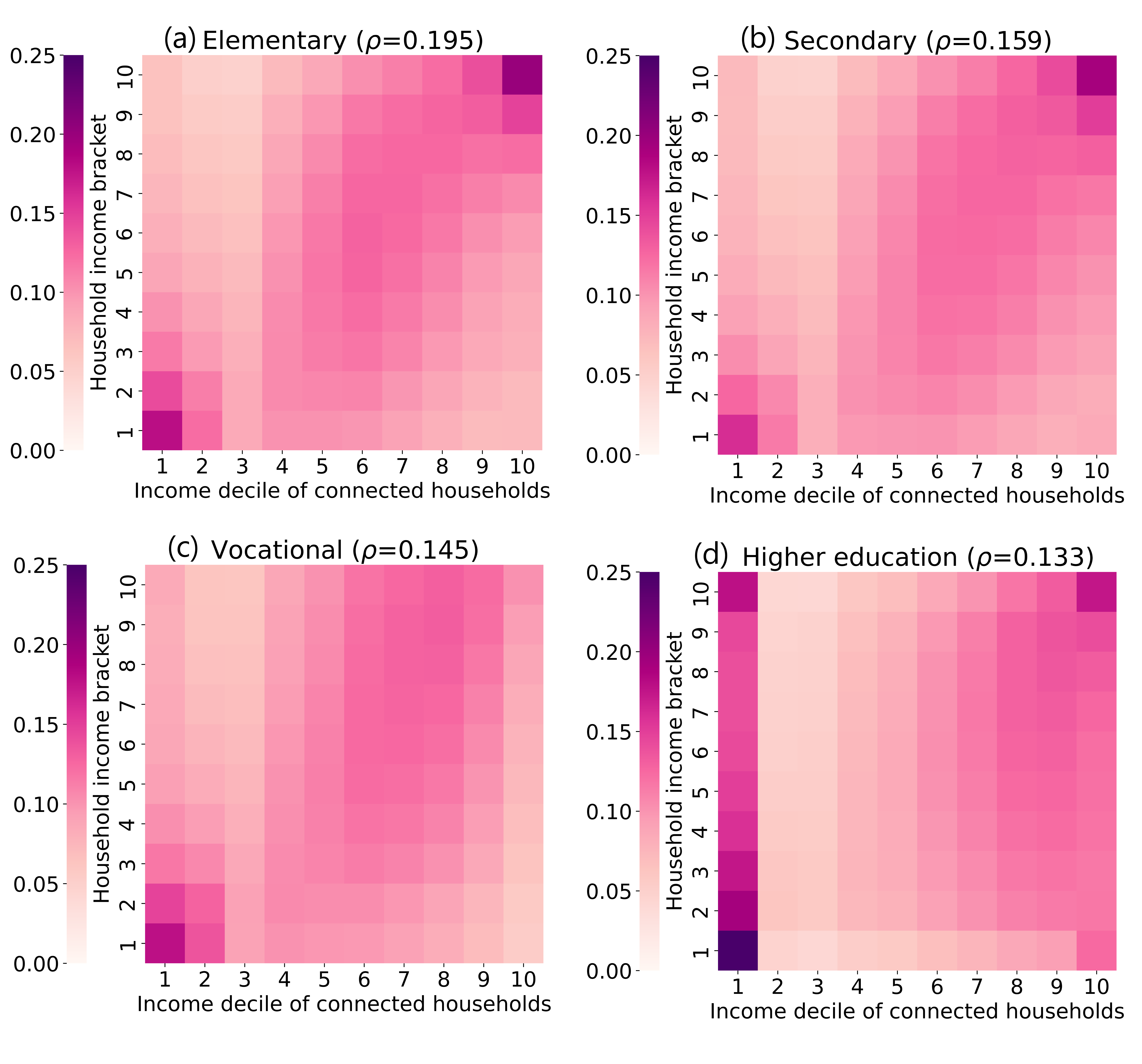}
\caption{\textbf{Mixing matrices for different stages of education:} (a) elementary education, (b) secondary education, (c) vocational institutions, (d) higher education.}
\label{fig:school_layers_combined}
\end{figure}

\renewcommand*{\bibfont}{\small}

\end{document}